# Recall Gabor Communication Theory and Joint Time-Frequency Analysis

Xiang-Gen Xia

University of Delaware

Newark, DE 19716, USA

Email: xianggen@udel.edu

**Abstract**

In this article, we first briefly recall Gabor's communication theory and then Gabor transform and expansion in signal processing, and also its connection with joint time-frequency analysis.

### 1. Gabor Communication Theory

In the 40s of the last century, there were two major concepts about digital communications. One is by Claude Shannon, i.e., the well-known Shannon communication theory published in the 1948 Bell System Tech Journal, which has become the foundation in modern digital communications. This is very well known these days and I will not elaborate any more about it here. The other is by Dennis Gabor who published "Theory of Communication" Part 1, Part 2, and Part 3 in 1946 [1]. In comparison, Shannon's theory is more from the statistical point of view and more mathematical, while Gabor's theory is more from structural or deterministic point of view and more physical, and also from more system view point. I will try to elaborate Gabor's communication theory a little more below.

In Gabor's communication theory, digital transmission is basically to tile the time and frequency plane. Given a time and frequency plane (or a channel), how many atoms (called Gabor atoms) can be packed and separated in the plane is kind of the physical capacity of the channel. The smaller the atoms are, the larger the capacity is, i.e., more different signals can be separated/detected. In other words, the smaller the product of the band width and time width is, the larger the capacity is. Unfortunately, due to Heisenberg's uncertainty principle, the minimum product of the band width and the time width is ½, which in fact provides the physical capacity for a given time and frequency plane, no matter single or multi-user communication systems. This is the fundamental limit in physics and unfortunately, these days many people have forgotten this fundamental physics limit. I think that a key difference with Shannon communication theory is that in Gabor's framework, the discrete signals $S_n$ and $S_{n,k}$ below are not necessarily quantized, i.e., not necessary in bits, and I think that it might be better to call Gabor's as semi-digital.

Since Gaussian pulse $p(t)= a \exp(-bt^2)$ for non-zero constants *a* and *b>0* reaches the lower bound ½ of the product of the band width and time width, Gaussian pulse is the most compact pulse (or the most time-frequency localized pulse) in the product domain of time and frequency. This is the reason why it is used in the following transmission signal

$$s(t) = \sum_n S_n p(t - nT) \tag{1}$$

where $S_n$ are the digital symbols to transmit and *T* is the time duration of a digital symbol to transmit. In fact, it is the reason why Gaussian pulse is used in the 2G GSM standard. The transmission signal in (1) is only in time domain. If it is tiled in both time and frequency domains, the signal becomes

$$s(t) = \sum_{k,n} S_{n,k} p(t - nT) \, \exp\left(j \frac{2\pi kt}{T}\right) \tag{2}$$

It is the formula (1.29) in [1] where *p(t)* is the Gaussian pulse. In other words, in terms of the time and frequency tilings, no pulse is better than the Gaussian pulse, no matter it is single or multi-user communication systems, for a given time and frequency rectangular window/plane.

In fact, the frequency shift 1/T in (2) can be generalized to a general *W* as

$$s(t) = \sum_{k,n} S_{n,k} p(t - nT) \, \exp(j2\pi kWt) \tag{3}$$

under the condition $TW \leq 1$, where the functions $p_{n,k}(t)=p(t-nT) \exp(j2\pi kW)$ for integers *n* and *k* are called Gabor atoms. The transmission signal (2) is a special case of (3) when *W=1/T*. If we sample time variable *t* in (3) and use the rectangular pulse, i.e., *p(t)* is the rectangular pulse of length *T*, for example, *t=l, T=N* and *W=1/N*, (3) is an OFDM signal without cyclic prefix (CP). If *p(t)* is a more general pulse, (3) is generalized OFDM (GFDM). In other words, the form of GFDM may be traced back to Gabor's 1946 paper [1].

In other words, using pulses in signal transmission appeared from the very beginning of digital communications by Gabor and it has been always natural to do so since then. However, it is often in-purposely skipped for convenience in theoretical studies.

## 2. Short-Time Fourier Transform, Gabor Expansion and Transform, and Joint Time-Frequency Analysis

The two dimensional signal $S_{n,k}$ in (3) can be thought of as a two dimensional sampling of the following short-time Fourier transform (STFT)

$$S(t,f) = \int s(\tau)\gamma(\tau - t)\exp(-j2\pi f\tau)d\tau \qquad (4)$$

where $\gamma(t)$ is called an analysis window in time domain. STFT is also called windowed Fourier transform. It was successfully applied in speech analysis in 1970's. Then, STFT was extensively studied in the era of wavelets of the 1990's as the first joint time-frequency analysis (JTFA) technique. Due to Gabor's contribution, the two dimensional sampling $S_{n,k}$ in (3) of the STFT $S(t,f)$ of signal $s(t)$ in (4):

$$S_{n,k} = \int s(\tau)\gamma(\tau - nT')\exp(-j2\pi kW'\tau)d\tau \qquad (5)$$

is called the Gabor transform of signal $s(t)$ and the pulse $p(t)$ in (3) is also called a synthesis window. The representation (3) is called Gabor representation (or expansion) of signal $s(t)$. The pair (3) and (5) are called Gabor representation and transform pair. More on the sampling distances on times $T$ and $T'$ and on frequences $W$ and $W'$ for (3) and (5) to hold simultaneously was studied thoroughly by Daubechies in [2]. For finite length discrete time signals, it is called discrete Gabor transform (DGT) and inverse DGT (IDGT) [5]. Due to the best time and frequency localization property of Gaussian window function/pulse, Gabor representation and transform have important applications in signal analysis, see for example [14].

For the relationship between the synthesis and analysis window functions $p(t)$ and $\gamma(t)$ for (3) and (5) to hold simultaneously either in continuous time or discrete time, Wexler-Raz [3] obtained an identity called Wexler-Raz identity in continuous time, discrete time, or finite length discrete time. The continuous time form of Wexler-Raz identity is

$$T_0 W_0 \int p(t)\gamma^*(t - mT_0)\exp(j2\pi W_0 t)dt = \delta(m)\delta(n) \text{ for all m and n}, \qquad (6)$$

where $T_0 = 1/W$ and $W_0 = 1/T$, and $T' = T$ and $W' = W$ in (5). Its discrete time versions are similar.

The relationship (6) corresponds to the biorthogonality, since the synthesis and the analysis windows $p(t)$ and $\gamma(t)$ may not be the same. The designs of these two window functions or pulses $p(t)$ and $\gamma(t)$ play an important role not only in signal analysis but also in communications when signal $s(t)$ is transmitted at transmitter and $S_{n,k}$ need to be detected at receiver.

In the oversampling case (for example, in the continuous time case, oversampling means *TW*<1, and in the finite length discrete time signal case, oversampling means that the number of points in the DGT domain is more than the signal length in time domain), for a given synthesis window *p*(*t*), the analysis window γ(*t*) satisfying the Wexler-Raz identity (6) is not unique. The optimal γ(*t*) (called the most-orthogonal-like, i.e., $\min\|\gamma(t) - p(t)\|$) was proposed in [4] and obtained in [5] for finite length discrete time signals and in [6] for more generally linear transformed most-orthogonal-like set-up (i.e., $\min\|\gamma(t) - Ap(t)\|$ where A is any linear operator with a certain condition) in either discrete or continuous time.

Interestingly, the generalized solution obtained in [6] is the same as the one without linear transformation, i.e., the same as that obtained in [5]. Also, interestingly, the optimal most-orthogonal-like solution coincides with the minimum norm solution in either discrete time [5] or continuous time [6], where these two solutions correspond to the two special cases of the above general optimization problem set-up in [6] when the linear operator *A* is the identity *I* and 0, respectively. For a finite discrete time, the DGT of a finite length signal of 1 dimension becomes a 2 dimensional matrix. Its matrix rank was investigated and determined in [7].

A JTFA is for non-stationary signal analysis and usually concentrates a signal (such as a chirp signal) while spreads noise, which means that the signal-to-noise ratio (SNR) in a JTFA domain may be increased over that in time domain. A quantitative SNR analysis in a JTFA domain is given in [8, 9, 10], which basically says that the SNR in a JTFA domain increases over the SNR in time domain linearly with the sampling rate in time domain of a signal. As we know that one important purpose for signal transformation is to clean a noisy signal, since the SNR may be increased in the transformation domain over that in time domain and it makes the cleaning easier in the transformation domain.

When a signal is non-stationary or wideband, the traditional filtering in the Fourier domain may not work well, because the Fourier transform may not be able to concentrate the signal, i.e., the SNR in frequency domain may not increase much over that in time domain. In this case, one may apply a JTFA that may concentrate the signal while it spreads the noise, i.e., the SNR in a JTFA domain may be increased significantly over that in time domain as mentioned above. Thus, one may apply a filtering in the JTFA domain.

However, since a JTFA is usually not an onto transform, after the filtering/clean-up in the JTFA domain, the filtered two dimensional signal may not correspond to a time domain signal, i.e., it may not be the JTFA of any time domain signal. Thus, filtering using JTFA (time-variant filtering) is not as trivial as that using the Fourier transform (time-invariant filtering). In this case, an iterative time-variant filter using DGT/IDGT is proposed in [11] with the convergence analysis. It says that the convergence is good for the most-orthogonal-like pair window functions. This DGT based time-variant filtering is used in system identification in low SNR environment in [12] by transmitting chirp signals, which may have applications in radar for target detection and in communications for channel estimation.

No transform can be universally good for every signal, including JTFA. In my opinion, JTFA works particularly well for chirp type signals including high order chirps. A discrete chirp-Fourier transform (DCFT) is proposed in [13] for linear-chirp signal matching, where it shows that DCFT works optimally when the signal length is a prime number. A good book on JTFA is [14] by Qian and Chen.

### 3. Conclusion

In this short article, we have briefly recalled Gabor communication theory in the 1940's, which is from a different perspective of Shannon's, and more from the physics and signal processing point of view, by packing signals in a time and frequency plane. In the later years, it has become Gabor atoms, Gabor transform, and Gabor expansion/representation for nonstationary signal analysis as a major JTFA.